\documentclass[]{tMOP2e}
\usepackage{graphicx}
\usepackage{natbib}

\begin{document}

\doi{10.1080/0950034YYxxxxxxxx}
 \issn{1362-3044}
\issnp{0950-0340} \jvol{00} \jnum{00} \jyear{2011} \jmonth{10 Febrary}

\markboth{Travis Horrom \textit{et al.}}{Quadrature noise in light propagating through a cold $^{87}$Rb atomic gas}

\articletype{Research paper}

\title{Quadrature noise in light propagating through a cold $^{87}$Rb atomic gas}

\author{
Travis Horrom$^a$,
Arturo Lezama$^b$,
Salim Balik$^c$, \\
Mark D.\ Havey$^c$,
and
Eugeniy E.\ Mikhailov$^a$ \\
$^a$ {\em{Department of Physics, The College of William and Mary, Williamsburg, VA 23187, USA}};
\\$^b$ {\em{Instituto de F\'isica, Universidad de la Rep\'ublica, Casilla de correo 30, 11000 , Montevideo, Uruguay}};
\\$^c$ {\em{Department of Physics, Old Dominion University, Norfolk, VA 23529, USA}};
\\\vspace{6pt} \received{\today}
}

\maketitle

\begin{abstract}

We report on  the study of the noise properties  of laser light propagating
through a cold $^{87}$Rb atomic sample  held in a magneto-optical trap. The
laser  is  tuned  around  the  $F_g=2 \to  F_e=1,2$  D$_1$  transitions  of
$^{87}$Rb. We  observe quadrature-dependent noise  in the light  signal, an
indication that it may be possible  to produce squeezed states of light. We
measure  the minimum  and maximum  phase-dependent noise  as a  function of
detuning  and  compare  these results to  theoretical predictions  to  explore  the  best
conditions for light squeezing using cold atomic Rb.

\begin{keywords}
    quantum fluctuations,
    squeezed states,
    polarization self-rotation,
    atomic noise
\end{keywords}
\end{abstract}

\section{Introduction}

The    electromagnetic   wave    quantum   operator    is   described    in
terms   of    two   quadrature    operators   $X_+=\frac{1}{2}(a^{\dag}+a)$
and  $X_-=\frac{i}{2}(a^{\dag}-a)$~\cite{scullybook,bachor_guide_2004}. The
Heisenberg  uncertainty  principle   sets  the  limit  on   how  small  the
fluctuations of  these quadratures can  be: $\Delta X_+\Delta  X_-\ge 1/4$.
The  case when  $\Delta X_+  = \Delta  X_- =  1/2$ is  called the  standard
quantum  limit  (SQL),  or  shot-noise limit.  Coherent  states  (typically
generated by lasers) and the vacuum are well known examples of field states
were the  SQL is  achieved. There  is currently much  effort to  reduce the
measurement noise  below this limit  with so called ``squeezed''  states of
light, where  the quantum fluctuations  of one of the  quadratures is reduced to
below the SQL~\cite{bachor_guide_2004}.

The  applications of  squeezed states  extend well beyond  precision measurements;
they were recently studied as a  carrier and probe for a quantum memory based
on atomic ensembles~\cite{akamatsu2004prl, appel_sq_quantum_memory_Rb_2007,
furusawa08_prl_sq_eit}. One  of the difficulties  for these studies  is the
lack of a  strong squeezing source at atomic  transition frequencies. While
nonlinear  crystal based  squeezers  can generate  an  impressive 11~dB  of
squeezing at 1064~nm~\cite{schnabel2010with11db}, they fail to deliver high
amounts of squeezing  at shorter wavelengths, since the  crystal windows of
transparency  lie  at higher  wavelengths.  So  far,  the record  value  of
squeezing at 795~nm is 5~db~\cite{hetet_squeezed_at_D1_Rb_2007}.

An      alternative     way      of     generating      squeezed     states
based       on       the       polarization       self-rotation       (PSR)
effect~\cite{budker2002rmp,     rochester_self-rotation_2001,     Davis:92,
PhysRev.137.A801,    novikova_large_sr_squeezing_2002}     was    suggested
in~\cite{matsko_vacuum_2002}. PSR  is a  nonlinear optical  effect observed
when  elliptically   polarized  light,  with  wavelength   tuned  near  an
atomic  transition,  experiences a  rotation  of  its polarization  ellipse
while  propagating  through  an  atomic  medium.  Since  the  intensity  of
the  left and right  circular  polarization  components are  different  in
elliptically polarized  light, this  leads to  unequal AC-Stark  shifts and
optical  pumping of  the  different atomic  Zeeman  sublevels resulting  in
circular birefringence.  Phase differences  in the  propagation of  the two
circular  components  of  the  light result  in  the  polarization  ellipse
rotation~\cite{novikova_ac-stark_2000,       rochester_self-rotation_2001}.
Unlike the Faraday effect, PSR is observed at zero magnetic field.

Several   research  groups~\cite{ries_experimental_2003,   mikhailov2008ol,
mikhailov2009jmo,  hsu_effect_2006, lezama_numerical_2008,  grangier2010oe}
have explored  the generation of squeezing  using the PSR effect  in hot Rb
vapor.  In all  of these  cases,  the amount  of squeezing  was about  1~dB
below  the SQL,  which  is  smaller  than the  original  prediction  of  Matsko
in~\cite{matsko_vacuum_2002}. This is attributed to excess atomic noise and
an inefficient  light-atom interaction  due to  Doppler broadening  in the hot  Rb
vapor samples used in the experiments.

Several groups have suggested  that in a cold atomic cloud,  the PSR effect will
yield higher  squeezing through the reduced thermal motion of  the interacting
atoms~\cite{mikhailov2009jmo,  hsu_effect_2006}.  Our  group  has  recently
reported the study  of the PSR effect  in a cold $^{87}$Rb cloud  held in a
magneto-optical trap (MOT)~\cite{mikhailov_psr_mot2011}.

In this paper we report on our theoretical  and experimental studies of the light
quantum  noise  modification  under  conditions  of  PSR  in  an  ultracold
$^{87}$Rb cloud. In the following sections, we first review our theoretical approach, and some results.  This is followed by description of the experimental apparatus and details pertinent to the measurements. After presentation and discussion of the results, we close with a summary and brief perspective on application of the PSR effect in an ultracold gas to generate squeezed states of light.

\section{Polarization self rotation squeezing theory}

According to  the original PSR  squeezing theory~\cite{matsko_vacuum_2002},
linearly  polarized  light slightly  detuned  from  a transition  resonance
of  an  atom  generates  squeezed vacuum  in  the  polarization  orthogonal
to  that  of  the  incident  light as  it  propagates  through  the  atomic
medium. The  amount of squeezing is  expected to increase in  proportion to
the  PSR effect.  However,  it  was quickly  understood  that the  original
treatment,  based on  the nonlinear  susceptibility of  the atomic  medium,
is  too  simplistic  since  it  does  not  account  for  the  excess  noise
introduced  by  the  atoms.  Although  atomic  absorption,  which  degrades
the  squeezing,   was  phenomenologically   taken  into   account,  sources
of  excess  noise,  such  as   amplified  spontaneous  emission,  were  not
considered~\cite{hsu_effect_2006}. However, excess noise  is present in all
experimental  observation  of  PSR  squeezing~\cite{ries_experimental_2003,
mikhailov2008ol, mikhailov2009jmo, grangier2010oe}. Under some experimental
conditions,  it  dominates  the   light  fluctuations  and  observation  of
squeezing becomes impossible~\cite{hsu_effect_2006}.

To  properly describe  the light  fluctuations after  interaction with  the
atomic  sample,  the quantum  fluctuations  of  the atomic  operators  need
to  be incorporated  into  the  treatment. This  can  be  achieved via  the
Heisenberg-Langevin  Equations that  incorporates  the atomic  fluctuations
through the use of stochastic forces.  Such an approach for the description
of PSR vacuum squeezing was  first attempted in \cite{hsu_effect_2006} on a
model four-level scheme and in \cite{lezama_numerical_2008} for a two level
system  including  the complete  Zeeman  degeneracy.  The first  successful
numerical  modeling  of  an  experimental  observation  of  PSR  squeezing
was  reported in  \cite{mikhailov2009jmo}.  There, it  was  shown that  the
complete excited state  hyperfine structure plays an essential  role in the
determination of the noise properties of  the light. In the present work we
have used  the numerical treatment used  in \cite{mikhailov2009jmo} applied
to  an ensemble  of cold  atoms. The  details of  this calculations  can be
found in  \cite{lezama_numerical_2008}. We briefly remind the reader of the essential
ingredients.

\begin{figure}[h!]
\includegraphics[width=0.35\columnwidth]{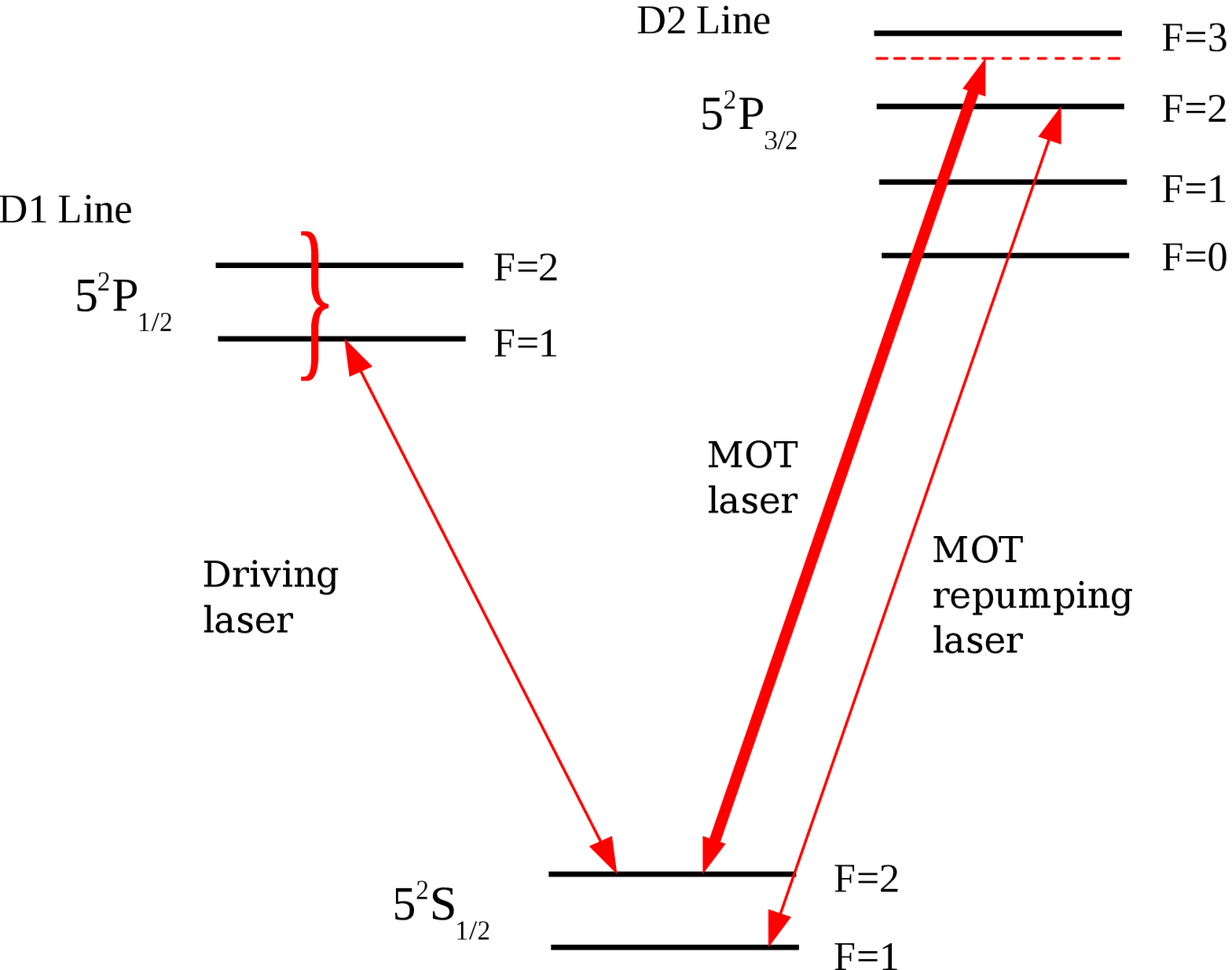}%
\includegraphics[width=0.6\columnwidth]{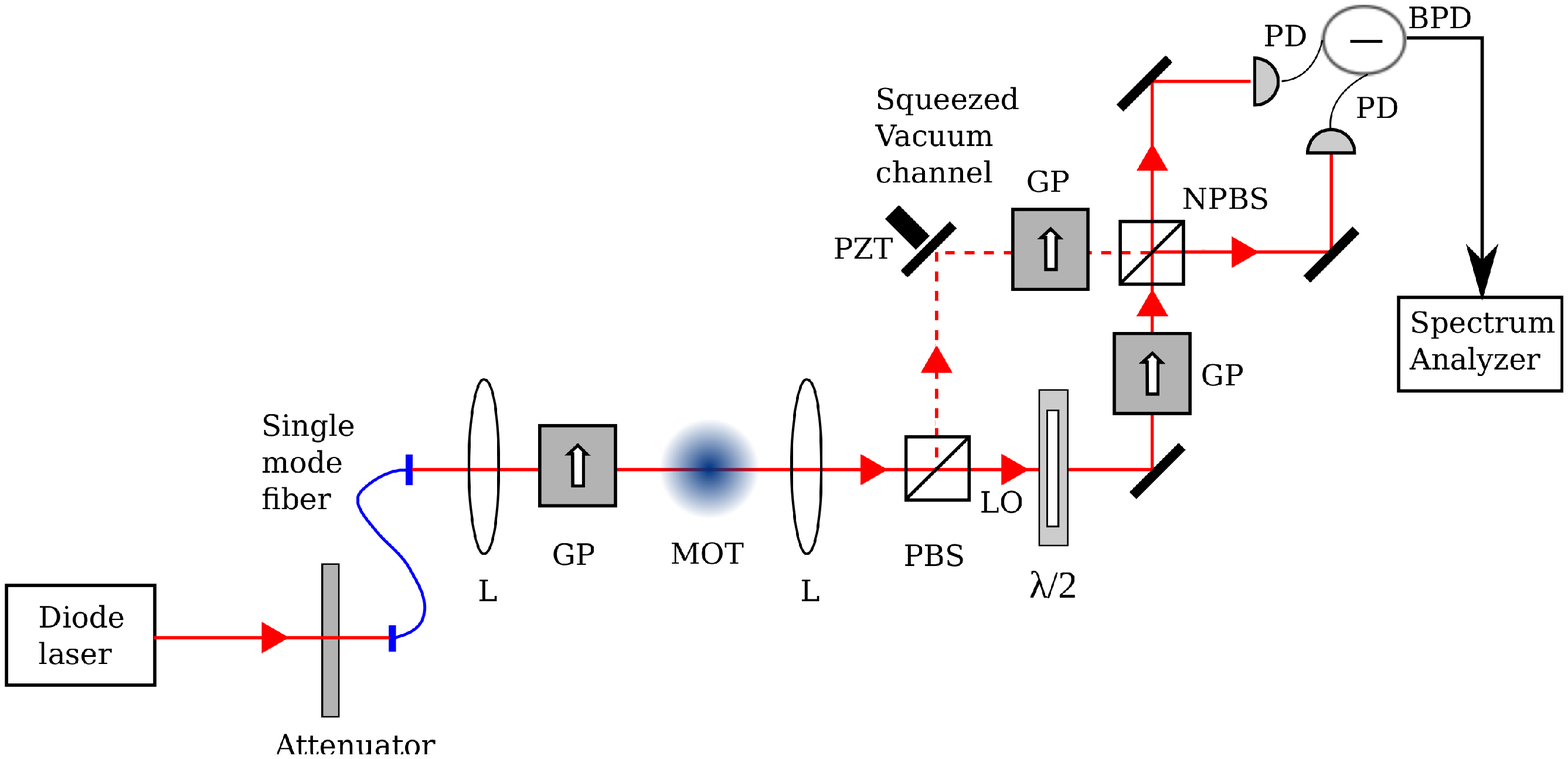}%
\caption{
    Left:
    Relevant laser fields on transition diagram showing D$_1$ and D$_2$ $^{87}$Rb lines.
        \label{fig:level_scheme}
    Right:
    Schematic diagram of the experimental setup.\label{fig:exp_setup}
    }
\end{figure}

Since  in   our  experiment  the   probe  laser  is  scanned   in  the vicinity
of  the $F_g=2   \to  F_e=1,2$  transitions   of the  $^{87}$Rb  D$_1$   line  (see
Fig.~\ref{fig:level_scheme}), we have taken  into account both relevant upper hyperfine states
of $^{87}$Rb ($F_e=1,2$).  The ground level  $F_g=1$ was neglected
since it  is detuned by 6.8~GHz  from the transition of interest.  The complete
Zeeman structure of  all three levels is considered. The  decay rate of the
upper  states is  $\Gamma$ ($\Gamma  =2\pi  \times 6$MHz)  and the  overall
phenomenological  decay  rate  for  atomic coherences  and  populations  is
$\gamma$  ($\gamma  \ll  \Gamma$).  The  ambient  static and spatially uniform magnetic  field  is  $B$.
The  incident  linearly  polarized  driving  field,  assumed  to  be  in  a
coherent  state,  has a  Rabi  frequency  $\Omega  =  \mu E/  \hbar$  where
$E$  is the  strength  of the  probe  light electric  field  and $\mu$  the
\emph{reduced} dipole moment matrix  element for the $5S_{1/2}\to 5P_{1/2}$
D$_1$ transition. The  atomic medium is characterized  by its cooperativity
parameter $C\equiv\frac{\eta  L \omega \mu^{2}}{2 \varepsilon_{0}  \Gamma c
\hbar}$ where $\eta$ is the atomic  density, $L$ the medium length. We note
that the  cooperativity parameter is  equal to $1/4$ of  the \emph{reduced}
resonant optical density of the medium.

The  Heisenberg-Langevin Equations  for  atoms and  fields are  numerically
solved at  steady state. For  this, the loss of  atoms at rate  $\gamma$ is
compensated  by  source  terms  representing the  arrival  of  fresh  atoms
isotropically  distributed  in  the   ground  state  Zeeman  sublevels.  As a consequence, the parameter $\gamma$ governs, at the same time, the decay of
coherence (in the absence of light) and the arrival of fresh atoms into the
system.

\begin{figure}%
\includegraphics[width=\columnwidth]{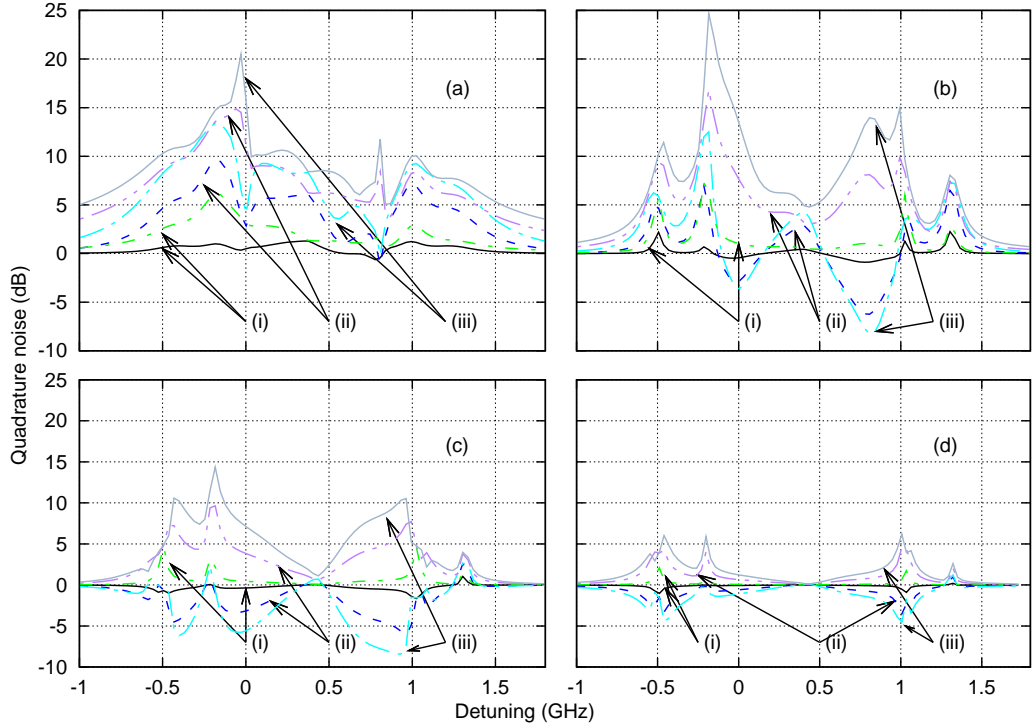}%
\caption{
    Phase-dependent noise vs. detuning for different cooperativity
    parameters and decay rates.
    Parameters are $\gamma=10^{-1}\Gamma$ (a), $\gamma=10^{-2}\Gamma$ (b),
    $\gamma=10^{-3}\Gamma$ (c), and $\gamma=10^{-4}\Gamma$ (d);  (i)
    C=$100$, (ii) C$=900$, (iii) C$=1700$; $\Omega =30\Gamma$, $B=0$ in all cases.
    \label{fig:noise_vs_detun_dif_coop}}
\end{figure}

The   results   of   our    numerical   calculations   are   presented   in
Fig.~\ref{fig:noise_vs_detun_dif_coop}. The  dependence of the  maximum and
minimum quadrature noise  levels in the output  field polarization component
perpendicular to that of the incident  driving field is shown as a function
of  the driving  laser  detuning  from the  $F_g=2  \to F_e=1$  transition.
Results for different  values of $C$ and $\gamma$ are  presented. It can be
seen that the  level of squeezing as well as the  contrast (difference between
maximum  and minimum  quadrature noise)  grows with  increased cooperativity
parameter (optical density). The contrast  diminishes as the laser detuning
increases. As expected,  the noise approaches the SQL noise  level for both
quadratures  for  a  very  far  detuned  laser,  since  in  this  case  the
light  does not  interact with the atoms.  Interestingly enough,  there is  an
optimum  in  the  transient  decay  rate  ($\gamma  \sim  10^{-2}$),  which
gives  the  highest  amount  of  noise  suppression  below  the  SQL.  This
may  seem  counter intuitive,  since  it  is  generally believed  that  PSR
squeezing is due to coherent effects  and should increase for longer ground
state  coherence  times. However,  the  ground  state coherence  time  also
depends on light  intensity through the off-resonance  optical pumping rate
$\alpha  \sim\frac{\Omega^{2}\Gamma}{\Delta^2}$  ($\Delta$ is  the  excited
states hyperfine levels separation).  Only when $\gamma\gtrsim \alpha$, the
decoherence  mechanisms represented  by $\gamma$  will limit  the squeezing
efficiency. On the other hand,  increasing values of $\gamma$ correspond to
larger number of ``fresh'' atoms  participating in the nonlinear interaction process resulting in larger modifications of the light fluctuations.

In the following  section we compare our experimental data  with the
results of the numerical predictions.

\section{Experimental Arrangement}

A schematic diagram of the  experimental apparatus is shown the right hand panel of Fig.~\ref{fig:exp_setup}.
In  order to reduce the  thermal energy of the atoms,  we use a
standard six beam  magneto optical trap (MOT) which is  described in detail
in~\cite{PhysRevA.79.033418}. An external cavity  diode laser, with a total
power  of $\approx  20$~mW,  detuned 18  MHz below  the  $F_g=2 \to  F_e=3$
$^{87}$Rb D$_2$ hyperfine transition is  used to create the trapping beams.
A  weaker  repumping  laser,  with  a total  power of $\approx  3$~  mW,  is
tuned  to the  $F_g=1  \to  F_e=2$ D$_2$  transition,  maintaining most  of
the  atomic population  in  the $5^2  S_{1/2}$  $F=2$  ground state.  See
Fig.~\ref{fig:exp_setup}  left panel  for a schematic diagram of  the atomic  energy
levels and  the applied laser  fields. Absorption imaging  of the  atomic cloud
shows that the MOT holds about $7\times10^7$ $^{87}$Rb atoms in a spherical
cloud with a  Gaussian distribution, with Gaussian radius  of about 0.5~mm.
Ballistic expansion  measurements indicate  that the atoms  are held  at an
average temperature of  300~$\mu$K and when the trapping  lasers are turned
off, the cloud  expands at a rate  of 200~$\frac{\mu \text{m}}{\text{ms}}$.
The trap magnetic field gradient has a  typical value of 5 G/cm. The sample
has a  peak density of  about $7\times10^9$ atoms/cm$^3$ giving  an optical
depth on the order of 2 for the driving laser transition.

Our linearly polarized driving laser,  with variable output power ($\mu$W -
$10$~mW),  was  tuned  around  the $F_g=2  \to  F_e=1,2$  transitions.  Our
previous  study  showed  that  if  the MOT  lasers  are  off,  the  typical
lifetime  of an  atom  in the  beam  is  less then  a  ms (especially  near
resonance)  due to  light pressure~\cite{mikhailov_psr_mot2011}  exerted by
the  driving  beam.  Since  we   wanted  to  perform  continuous  squeezing
experiments,  we keep  the MOT  lasers  on to  continuously repopulate  the
cloud  with Rb  atoms.  The presence  of  the MOT  trapping  beam helps  to
maintain  the  atomic  cloud  centered  around the  zero  of  the  magnetic
field.  In  previous  observations,  where  the  trapping  beams
were  turned  off,   we  could  observe  that  light   pressure  and  cloud
expansion resulted  in a  nonzero average  magnetic field  of the  order of
$B=10^{-5}$~Tesla~\cite{mikhailov_psr_mot2011}.

\begin{figure}%
\includegraphics[width=.5\columnwidth]{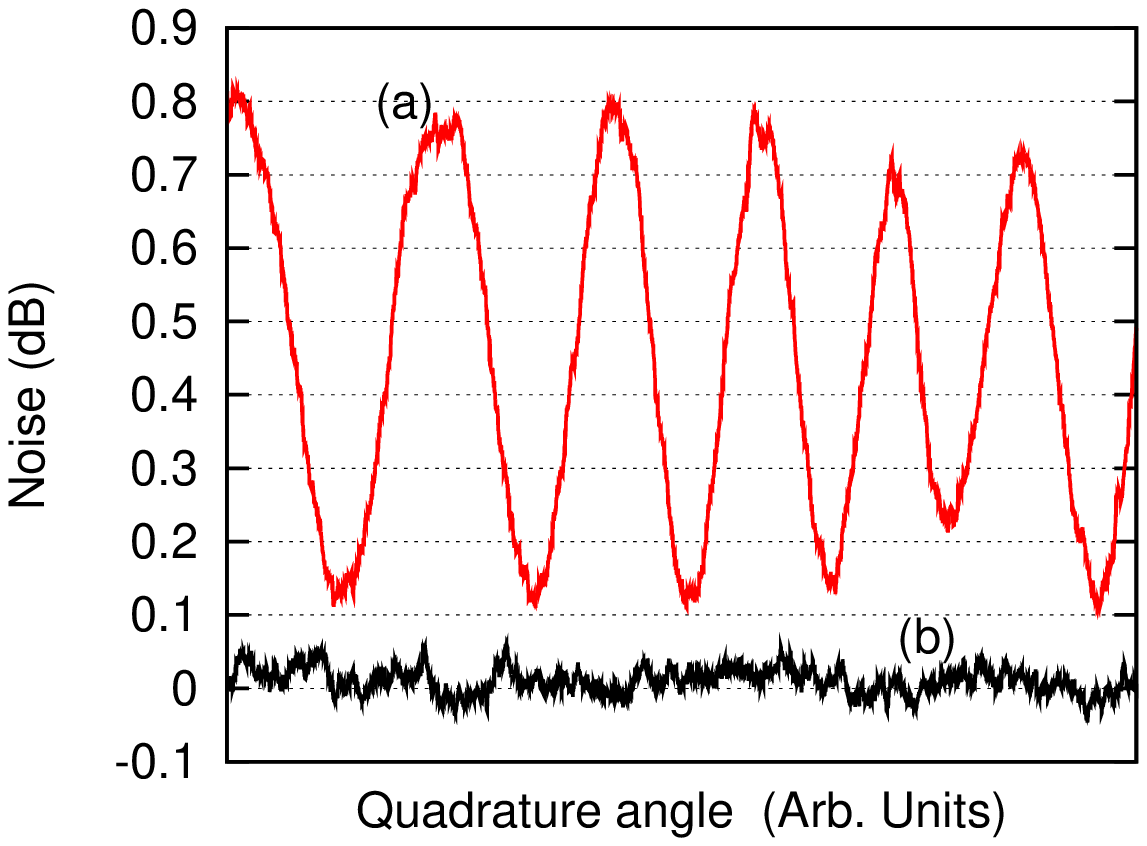}%
\includegraphics[width=.5\columnwidth]{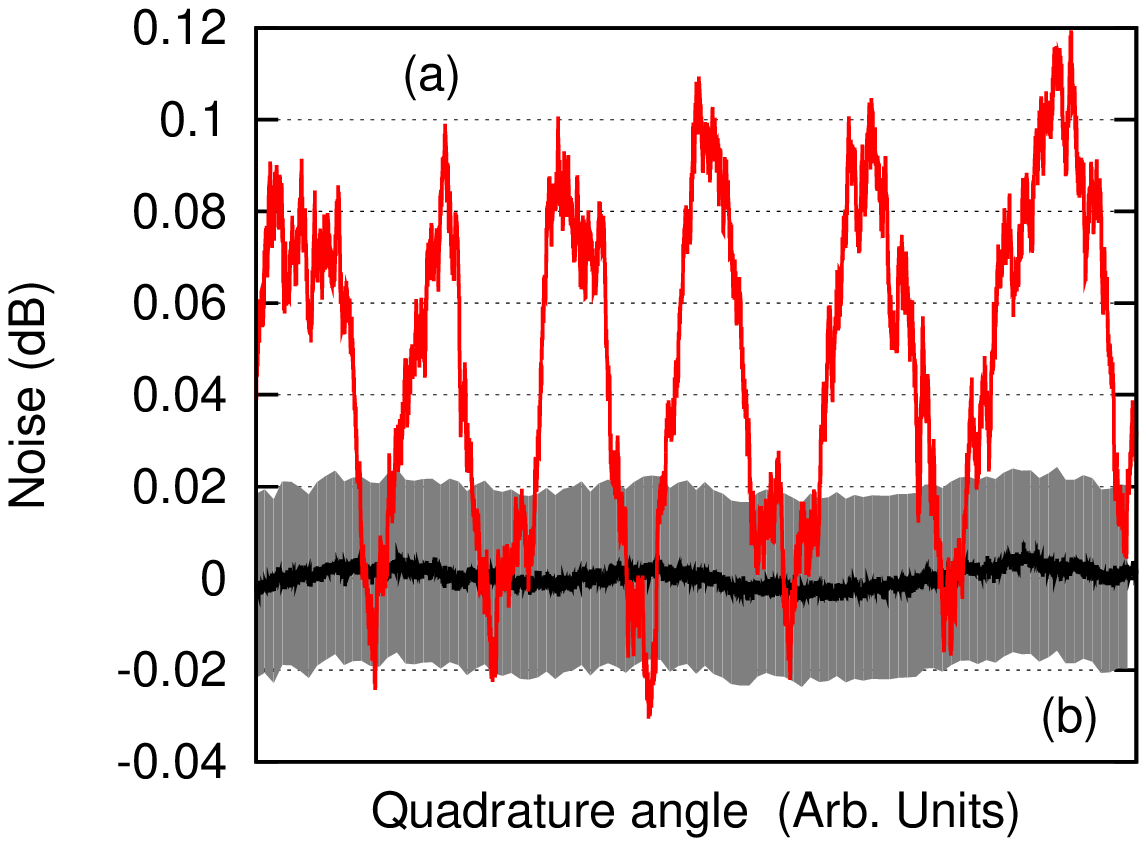}%
\caption{
    Left: Typical noise power dependence in the squeezed channel  versus the quadrature angle.
    Phase-dependent excess noise: Laser power $= 1.3$~mW, Detuning $= -200$~MHz
    Right: Experimental data with the highest observed degree of squeezing.
    Laser power $= 6$~mW, Detuning $= -220$~MHz.
    (a) modified quantum noise in the vacuum channel,
    (b) shot-noise level (shown with uncertainty band on the right).
    These noise traces are measured at
    1.4~MHz central frequency of the SA, RBW$=100$~kHz,
    and are averaged over 512 traces.
    \label{fig:noise_vs_quad_angle}%
}
\end{figure}

The driving laser light travels through a single mode fiber to achieve a
spatially clean  Gaussian beam,  and then passes  through a  Glan polarizer
(GP) to enforce a linear polarization of the driving beam. We use a pair of
lenses (L) to focus  the driving laser in the interaction  region to a beam
diameter of  around 250 $\mu$m  ($1/e$ intensity level). The Rb  cloud is
larger  than the  beam diameter  and serves  as a  reservoir of  cold atoms
during  the experiment.  We  then separate  the  linearly polarized  strong
driving  field from  the  squeezed vacuum  with  a polarizing  beamsplitter
(PBS). The strong port is attenuated to 100 $\mu W$ and serves as a local
oscillator (LO)  in the custom made  balanced homodyne detector (BPD).  The LO
field is  rotated by an extra  90$^\circ$ with a  half-wave plate in  order to
match the polarization of the vacuum  channel. The vacuum channel  and
LO fields pass through Glan polarizers in  order to improve the extinction ratio
of the PBS, and are finally mixed on a non-polarizing beam splitter (NPBS). The
beamsplitter outputs are directed to two matched  photodiodes (Hamamatsu S5106), each having a 93\% quantum  efficiency, where the two photo  currents are electronically subtracted.
We analyze the  remaining noise with a spectrum analyzer  at 1.4~MHz with a
resolution  bandwidth  (RBW)  of  100~kHz. The  overall  mode  matching  of
the  vacuum channel  to  the LO  mode  is checked  via  observation of  the
interference fringes with  visibility higher than 95\%.  The vacuum channel
has  a piezo  ceramic  transducer (PZT)  attached to  one  of the  mirrors.
This  allows us  to sweep  the  relative phase  between the  LO and  vacuum
channel, and consequently to measure  the noise in the different quadrature
projections.  Fig.~\ref{fig:noise_vs_quad_angle} shows  examples of  such a
sweep.  The 0~dB  noise  level  corresponds to  the  shot  noise, which  we
determine by introducing a solid block into the vacuum channel. Noise below
0~dB indicates squeezing. We note that  overall stability of the shot noise
level is  about $\pm 0.2$~dB, which  is governed by the  fluctuations of the LO
power and stability of the spectrum analyzer.

\section{Experimental Results}

Due  to  the relatively small  number  of  atoms ($\approx  10^5$)  interacting  with  the
PSR  driving  beam  the  overall   noise  contrast  is  below  0.8~dB  (see
Fig.\ref{fig:noise_vs_quad_angle}, left  panel,  for  the  highest  contrast
case), which  is significantly smaller in comparison with the PSR  squeezing contrast
typically observed in hot Rb cells. It is important to note in this comparison that the number of interacting  atoms in the
hot  cell is approximately one thousand  times higher.  The  experimentally observed  optical
density  is  only around  2,  which  is  far  from the  high  cooperativity
parameters  required to  achieve the  significant noise  contrast presented  in
Fig.~\ref{fig:noise_vs_detun_dif_coop}.

\begin{figure}
\centering
    \includegraphics[width=\columnwidth]{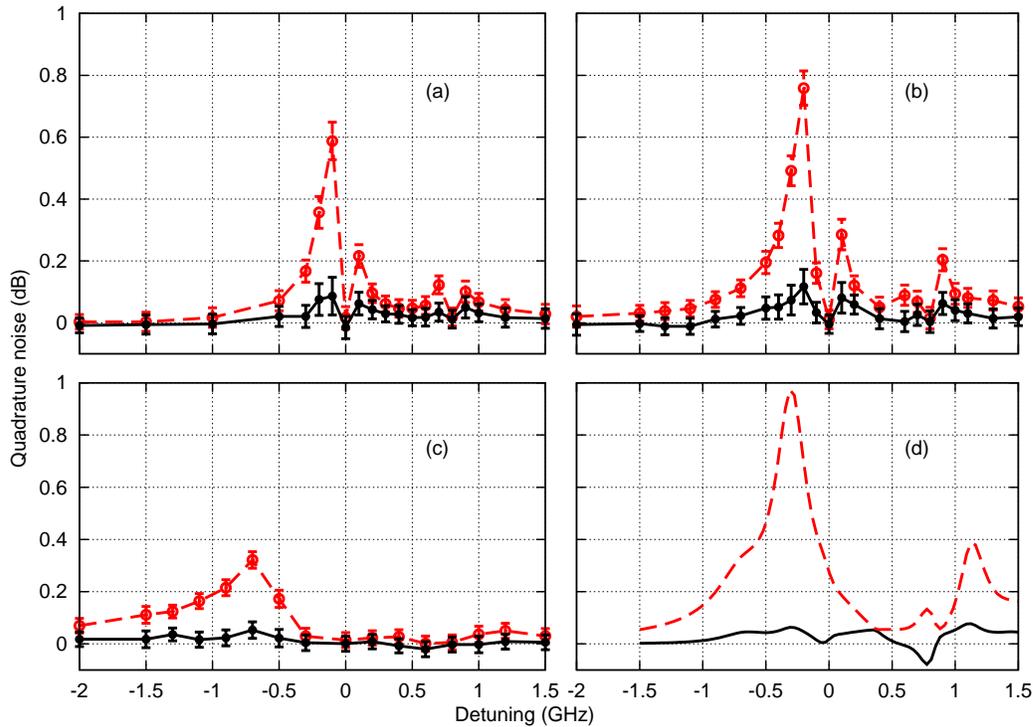}
\caption{
    Results of the experiment (a, b, and c) and numerical simulations (d)
    for
    minimum (solid line) and maximum (dashed line) noise levels dependence
    on the PSR driving laser detuning for
    different PSR driving laser powers.
    (a) laser power $0.47$~mW, (b) $1.3$~mW, (c) $7.5$~mW. Parameters for
    numerical simulation (d) power $=10$~mW, beam crossection
    $10^{-3}$~cm$^2$, $\gamma = 0.1\Gamma$, C$=10$, $B=0$.
    \label{fig:noise_vs_detun_exp_and_theory}
    }
\end{figure}

In  order  to  record  the minimum and maximum   noise  level  dependence  on  the  driving
laser  detuning,  we  set  our   laser  to  a  given  detuning  (controlled
with  a   commercial  wave   meter  with   10~MHz  accuracy),   and  recorded
noise  versus  the quadrature  angle  dependence   similar  to  that  shown  in
Fig.~\ref{fig:noise_vs_quad_angle}. On such a trace,  we note the maximum and
minimum noise levels, which provide two data points for each detuning shown
in  Fig.\ref{fig:noise_vs_detun_exp_and_theory}(a,b, and  c). Note  that at
detunings exactly matching the atomic  transitions (0 and 815~MHz), we have
zero noise  contrast and the  overall noise level  drops to shot  noise. We
attribute this  to the  strong light  pressure of the  driving beam  on the
atoms  at frequencies  very  close  to the  transitions,  which blows  away
the  atomic  cloud.  We  take  such contrast measurements versus  detuning  spectra  at
several  driving  laser  powers  and  see  that  contrast  initially  grows
with  power since  the nonlinear  PSR interaction  is increases  with laser
power,  but  then the  contrast  decreases  due  to  a stronger  effect  of
light  pushing the  atoms away from the interaction region with  increased power.  We  also  note that  the
highest contrast position  moves away from the  transition frequencies with
increasing  power (more  negative  for $F_g=2  \to  F_e=1$ transition,  and
positive for $F_g=2  \to F_e=2$) due to power broadening  of the transition
resonance.  This  effect is  often  seen  in  the  PSR squeezing  with  hot
Rb~\cite{grangier2010oe}. The theoretical predictions  of the noise spectra
match  the shapes  of  the  experimental traces  quite  well,  as shown  in
Fig.~\ref{fig:noise_vs_detun_exp_and_theory}d. In  this simulation  we have
taken  $\gamma  =0.1 \Gamma$  and  $B=0$.  The  relatively large  value  of
$\gamma$  was  chosen to  account  for  the  fact  that the  atomic  ground
state coherence  is strongly  perturbed by the  MOT trapping  and repumping
beams.  The  ambient  magnetic  field  in the  MOT  region  is  known  from
~\cite{mikhailov_psr_mot2011} to be on the order of  $B=0.01 \Gamma$ (in units of
the corresponding Zeeman frequency shift). Since $B \ll\gamma$ the magnetic
field influence  is negligible  and the  zero magnetic  field approximation is
justified.

\section{Summary and outlook}

We have observed overall quantum  noise modification  via the PSR effect  in an ultracold
$^{87}$Rb  atomic medium,  which is in  a  good agreement  with our  numerical
simulations.

We do not have compelling results showing squeezing below the SQL.  Our results are limited by the levels of excess noise, which are relatively small and the predicted squeezing
is on the order of a tenth of a dB, at the limit of our current resolution.
We  do however  see clear  phase-dependent excess  noise and,  depending on
conditions,  several points  where the  minimum  noise level  is very  near
shot  noise and  may  be  squeezed (see  Fig.~\ref{fig:noise_vs_quad_angle}
right  and  traces   in  Fig.~\ref{fig:noise_vs_detun_exp_and_theory}b  and
Fig.~\ref{fig:noise_vs_detun_exp_and_theory}c).  We attribute  the lack  of
obvious  squeezing to  a  low  number of  atoms  interacting  with the  PSR
driving  beam  in our  current  cold atom arrangement.  We  believe that  an instrument
with  a higher  optical density  will  result in  stronger squeezing with
noise below the SQL,  as  predicted by our  numerical  simulations.
Such  cold atom instruments are well within  the  experimental reach  of  the  current state  of
technology;  for instance, it is  possible  to have  a  large magneto optical trap  with  up to  $10^{10}$
atoms~\cite{1402-4896-81-2-025301}, or to create an asymmetric cigar-shaped
MOT  so  that  the  longer  dimension  can  be  aligned  with  the  PSR  driving
beam~\cite{Greenberg:07, Lin:08} thus achieving a quite substantial optical depth and a considerably higher number of interacting atoms.

With these  improvements and more  studies, we  believe that
generation of squeezed vacuum with  higher levels of noise suppression than
seen in hot vapor cells is achievable using PSR in cold atoms.


\section{Acknowledgments}

E.M. and  T.H. thank the support  of NSF Grant PHY-0758010.  A.L. wishes to
thank support from  ANII, CSIC, PEDECIBA (Uruguayan agencies),  and the APS
International Travel Grant  Program. Partial support of this  work (S.B and
M.H.) was  provided by  NSF Grant  PHY-0654226. Numerical  simulations were
carried in the SciClone Cluster (College of William and Mary).

\bibliographystyle{tMOP}

\end{document}